\documentclass[12pt]{article}

\usepackage{amsfonts}
\usepackage{amssymb}
\usepackage{epsfig}
\usepackage{rotating}
\usepackage{lineno}
\usepackage{color}
\usepackage{amsmath}
\usepackage{mathrsfs}
\usepackage{hyperref}
\usepackage{setspace}
\usepackage{grffile}
\usepackage{multirow}
\usepackage{verbatim}   
\usepackage{url}
 \usepackage{indentfirst}

\newcommand{\citep}{\cite}
\makeatletter
\def\url@leostyle{%
  \@ifundefined{selectfont}{\def\UrlFont{\sf}}{\def\UrlFont{\small\ttfamily}}}
\makeatother
\newcommand{\ck}[1]{{\bfseries\color{red}{$\langle$CHECK$\rangle$}}}
\newcommand{\cn}[1]{{\bfseries\color{red}{$\langle$Citation needed$\rangle$}}}

\begin{document}


\title{Patterns of Influenza Vaccination Coverage in the United States from 2009 to 2015}

\author{
 Alice P.Y. Chiu$^{1}$,  Duo Yu$^{1,2}$, Jonathan Dushoff$^{3,4}$ and Daihai He$^{1, \star}$
\vspace{0.2cm}\\
{\footnotesize $^1$ Department of Applied Mathematics, Hong Kong Polytechnic University, Hong Kong (SAR) China}\\
{\footnotesize $^2$ Department of Biostatistics, University of Texas Health Science Center at Houston, United States}\\
{\footnotesize $^3$ Department of Biology, McMaster University, Hamilton, ON, Canada}\\
{\footnotesize $^4$ M.G. DeGroote Institute for Infectious Disease Research, McMaster University, Hamilton, ON, Canada}\\
{\footnotesize $\star$ hedaihai@gmail.com}
}

\maketitle
\parindent 1cm

\section*{Abstract}
\subsection*{Background}
Globally, influenza is a major cause of morbidity, hospitalization and mortality.
Influenza vaccination has shown substantial protective effectiveness in the United States.
We investigated state-level patterns of coverage rates of seasonal and pandemic influenza vaccination, among the overall population in the U.S. and specifically among children and the elderly, from 2009/10 to 2014/15, and associations with ecological factors.

\subsection*{Methods and Findings}

We obtained state-level influenza vaccination coverage rates from national surveys, and state-level socio-demographic and health data from a variety of sources.
We employed a retrospective ecological study design, and used mixed-model regression to determine the levels of ecological association of the state-level vaccinations rates with these factors, both with and without region as a factor for the three populations.
We found that health-care access is positively and significantly associated with mean influenza vaccination coverage rates across all populations and models.
We also found that prevalence of asthma in adults are negatively and significantly associated with mean influenza vaccination coverage rates in the elderly populations.

\subsection*{Conclusions}

Health-care access has a robust, positive association with state-level vaccination rates across different populations.
This highlights a potential population-level advantage of expanding health-care access.

\section*{Introduction}\label{S:intro}

Globally, influenza is the cause of three to five million cases of severe influenza illness and 250,000 to 500,000 deaths annually \cite{WHO}. Influenza vaccine coverage rates in the United States are the highest among the Americas \cite{Palache}. The US Healthy People 2020 initiative set a target of 70\% vaccination coverage of both children and adults by 2020 \cite{HealthyPeople}. Vaccine effectiveness against seasonal influenza A and B virus infection and pandemic influenza (A(H1N1)pdm09) has demonstrated moderate effectiveness \cite{MMWR}.

Previous studies have investigated individual-level correlates of seasonal and pandemic influenza vaccination coverage rates in the US. It was found that younger age \cite{Linn,Galarce,Takayama}, lower levels of education \cite{Linn,Takayama}, lower household income \cite{Linn,Fox}, urbanicity\cite{Galarce}, having a history of chronic conditions or poor physical health\cite{Linn,Takayama}, lack of health insurance\cite{Takayama,Fox}, and smoking\cite{Takayama} were factors associated with lower vaccination coverage rates. In addition, non-Hispanic Black race was associated with lower rates of pandemic influenza vaccination coverage \cite{Uscher-Pines}.  However, whether these factors could explain the state-level variations have not been studied.

In the US, the Center for Diseases Control and Prevention (CDC) recommended seasonal influenza vaccination to all persons aged 6 months or older in 2010/11 following the 2009 influenza A (H1N1) pandemic \cite{CDC}. In this study, our aim is to identify how socio-demographic and health-related factors are ecologically associated with the mean state-level rates of influenza vaccination coverage among the overall population (i.e. persons aged six months or above) from 2009/10 to 2014/15.
Our secondary aim is to compare the association of these factors in the presence or absence of region as a factor.
In addition to our main analysis in the whole population, we studied these associations separately in the elderly (aged 65 or above) and children (aged 6 months to 18 years) sub-populations.

\begin{figure}[ht!]
\centerline{\includegraphics[width=17cm]{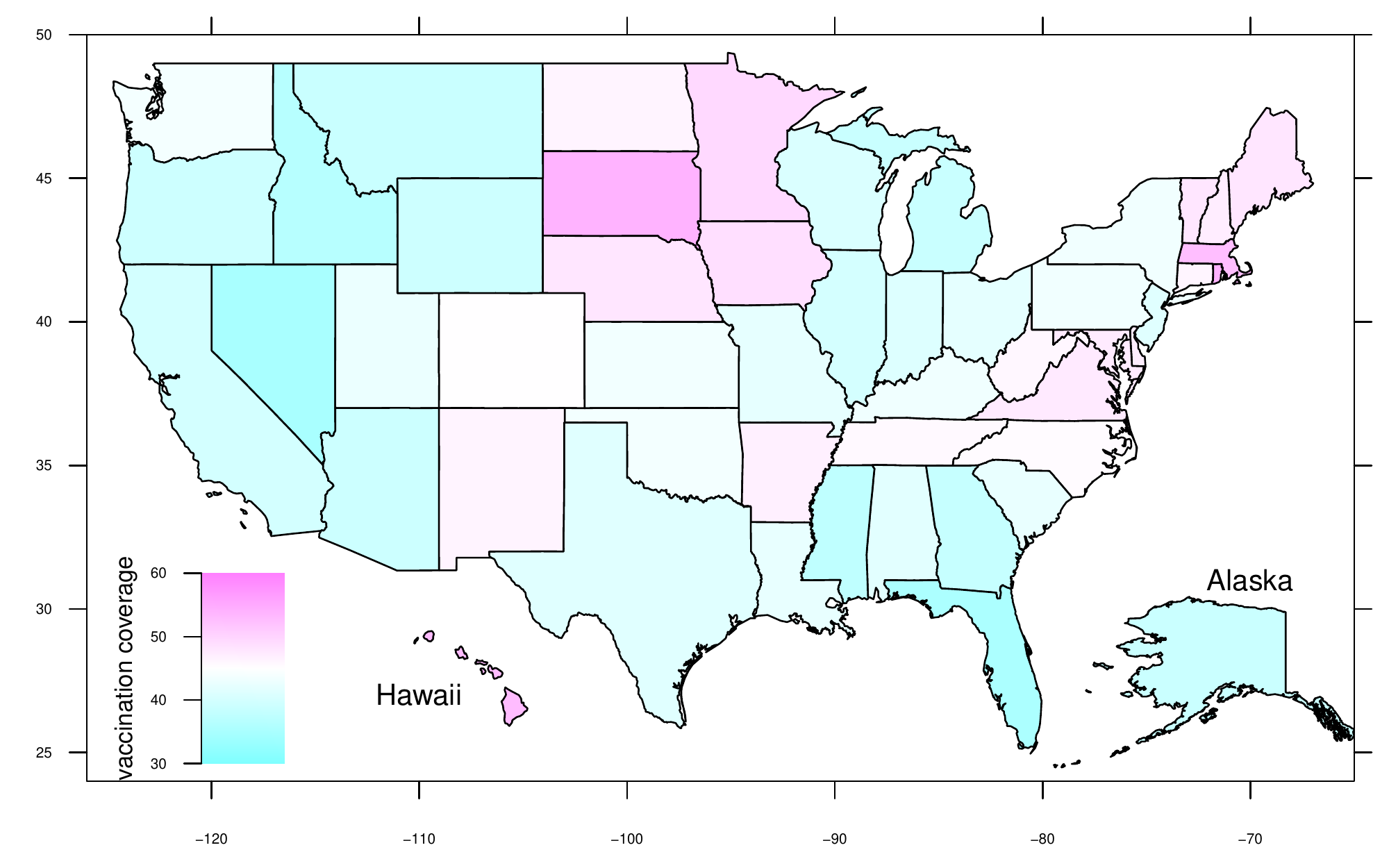}}
\caption{Mean influenza vaccination coverage rates of 50 States and DC.
Hawaii and Alaska are shifted, and Alaska is scaled by 1/3.}
\label{Fig:vacc}
\end{figure}

\section*{Methods}
\subsection*{Study Population and Setting}
Our study population (``overall population") consisted of all persons aged six months or older from 50 states and the District of Columbia in the U.S. from 2009/10 to 2014/15 who were eligible for influenza vaccination.
We also studied the elderly and children sub-populations within the overall population.
All data were obtained from publicly available sources.
Ethical approval was not required.

\subsection*{Study Design}
We adopted a retrospective ecological study design to assess the overall socio-demographic, health-related and regional correlates with the mean state-level vaccination coverage rates across the time period.
This study design does not permit the analyses of individual-level associations.
However, it is appropriate for studying the state-level vaccination coverage rates and their potential associations with these factors.

\noindent{\bf Influenza Vaccination Coverage Data}
\\
We used published data from Behavioral Risk Factor Surveillance System (BRFSS) \cite{BRFSS} and National Immunization Survey (NIS) \cite{NIS}.
These sources give influenza coverage rates for the 50 states plus the District of Columbia for each influenza seasonal year (from June through May) for the period from 2009/10 to 2014/15. BRFSS's study population included adults aged 18 year or above whereas NIS study population covered children aged 6 months to 17 years old. For 2009/10, due to the H1N1 pandemic, they report coverage for: seasonal vaccination, pandemic H1N1 vaccination, or at least one influenza vaccination. We used the last value.
These vaccination coverage rates ranged from 30\% to 60\%. We also computed the mean rates based on their annual rates from 2009/10 to 2014/15. These mean state-level rates are then used for further analyses.

\noindent{\bf Region} \\
We aggregated the 50 states and the District of Columbia into nine regions according to the US Census Bureau classification\cite{Region}.

\noindent{\bf Socio-demographic and Health-related Data}\\
To investigate factors associated with vaccination coverage rates, we searched publicly available sources for socio-demographic and health-related data.
Specifically, we collected secondary data for the following state-level variables in which their data definitions and data sources are described in more details below.

\noindent{\bf Health Care Access}\\
We obtained estimates from the CDC's Behavioral Risk Factor Surveillance System (BRFSS) Surveys in 2013. Health care access is defined as having any kind of health insurance, prepaid plans, government plans or Indian Health Service \cite{Access}.

\noindent{\bf Population Proportion of Children and Elderly}\\
We obtained population estimates from the US Census Bureau in 2013 to determine the proportion of children and elderly in each state \cite{AgeandRace}.

\noindent{\bf Per capita Personal Income Level}\\
We also obtained the per capita income level in 2013 from the US Bureau of Economic Analysis in each state \cite{Income}.

\noindent{\bf Educational Attainment}\\
Data about educational attainment is obtained from the US Census Bureau in 2013. It is defined as the proportion of population who had attained college degree or above \cite{Education}.

\noindent{\bf Racial or Ethnic Origin}\\
We defined race and ethnic origin into two variables: Hispanic and Black.
We classified persons into Hispanic or non-Hispanic origin.
We also defined persons as Black or non-Black (i.e.
White, Asian American, Native American or others, including persons who are multi-racial) \cite{AgeandRace}.

\noindent{\bf Prevalence of Asthma}\\
We estimated prevalence of asthma among the adult population using responded from the 2013 BRFSS survey \cite{Asthma}.

\noindent{\bf Missing Data}\\
There are missing data in the 2013/14 influenza vaccination coverage for two states. Since vaccination levels are relatively constant from year to year, we simply calculate mean coverage for those two states using the remaining years.
We filled the ``N/A" entry of ``Black" for  Wyoming using 1 minus the proportions of White, Hispanic, and Others, and yield 0.009845.

\noindent{\bf Statistical Analyses}\\
We determine the levels of ecological association of the state-level vaccination coverage rates with socio-demographic and health-related factors across the period from 2009/10 to 2014/15. We compared the levels of ecological association with and without region as a factor.
We conducted analyses for the overall population as well as the elderly and children sub-populations.
We developed six different models.
Model 1 is a linear model with socio-demographic and health-related variables as independent variables, and the mean vaccination influenza coverage rates of the overall population as dependent variable.
Model 2 is a linear mixed effects model with socio-demographic and health-related variables as fixed factors, and region as a random factor, and the mean vaccination coverage rates of the overall population as dependent variable.
Model 3 and Model 4 are similar to Model 1 and Model 2, except that mean vaccination coverage rates of the elderly populations are used as dependent variables.
For Model 5 and Model 6, they are also similar to Model 1 and Model 2, except this time we used the mean vaccination coverage rates of the children population are used as dependent variables.

In all six models, we standardised all variables to mean zero and unit standard deviation before performing analyses.
We conducted all statistical analyses using R (version 3.2.2). Statistical significance is assessed at the 0.05 level.

\section*{Results}
In Figure 1, we colored the map of the U.S. according to the mean vaccination rates.
Northwestern and Western regions displayed lower vaccination rates, while Mid-Western and Northeastern regions displayed higher rates.
Also, there were substantial state-level variations within Mid-western region.

We show regression estimates from Models 1 to 6 in Figures 2, S1 and S2.
The thick lines show 50\% confidence interval while the
thin lines show 95\% confidence intervals.
Figure 2 left panel depicted the regression estimates of the socio-demographic factors and health-related factors on the mean influenza vaccination coverage rates of the overall population (Model 1). It was found that health care access is the only factor that is positively and significantly associated with influenza vaccination rates at 5\% level.
Factors positively associated with the influenza vaccination rates are per capita income level, proportion of elderly, proportion of children and proportion of Hispanic.
Factors negative associated with the influenza vaccination rates are educational attainment, prevalence of  among adults and the proportion of Black.

In Figure 2 right panel, we depicted the  regression estimates of the linear mixed effect models of the mean overall population's influenza vaccination coverage rates(Model 2). Compared with Model 1, we have included region as a random factor.  All regression estimates are similar, and there were no changes in sign.

\begin{figure}[h!]
\includegraphics[width=8cm]{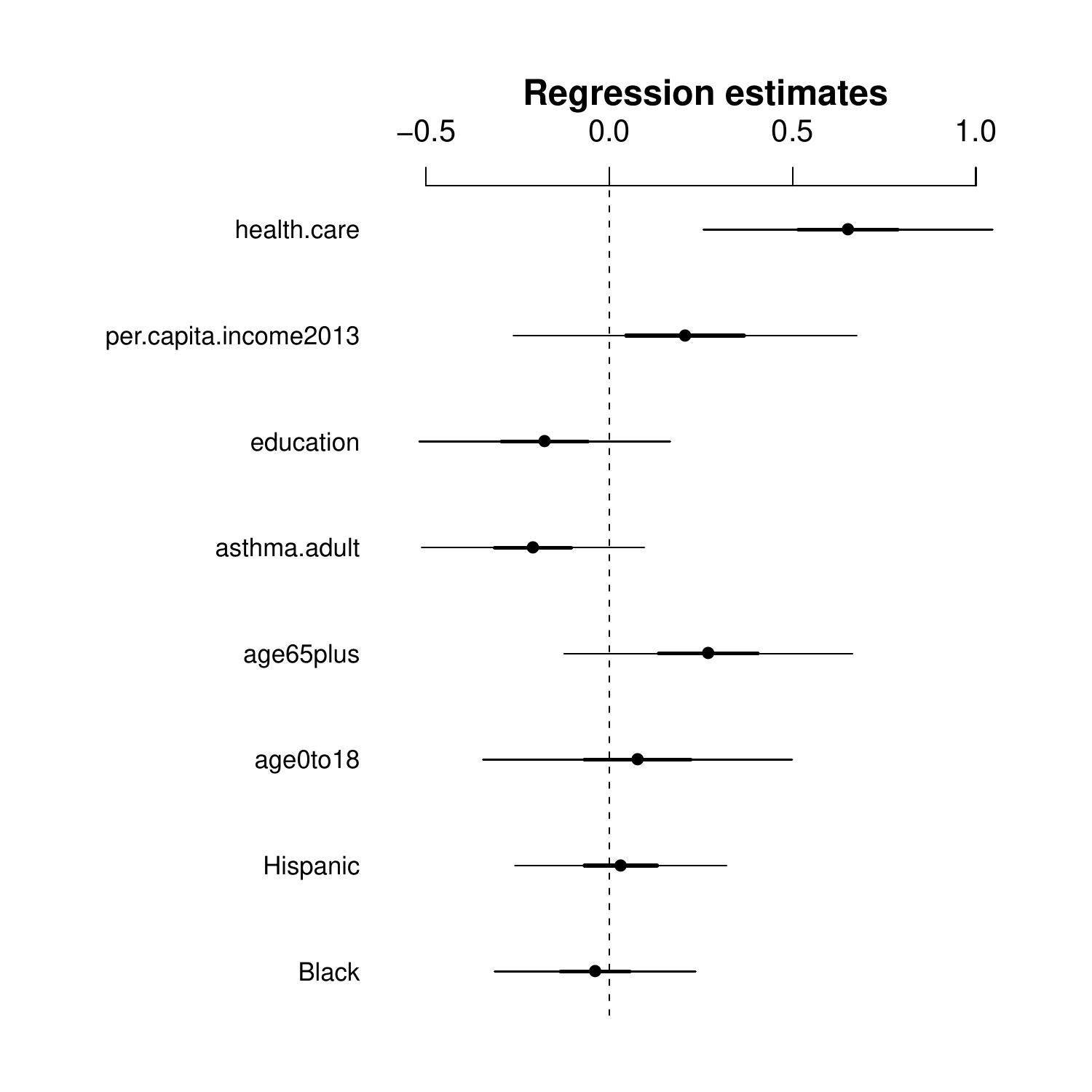}
\includegraphics[width=8cm]{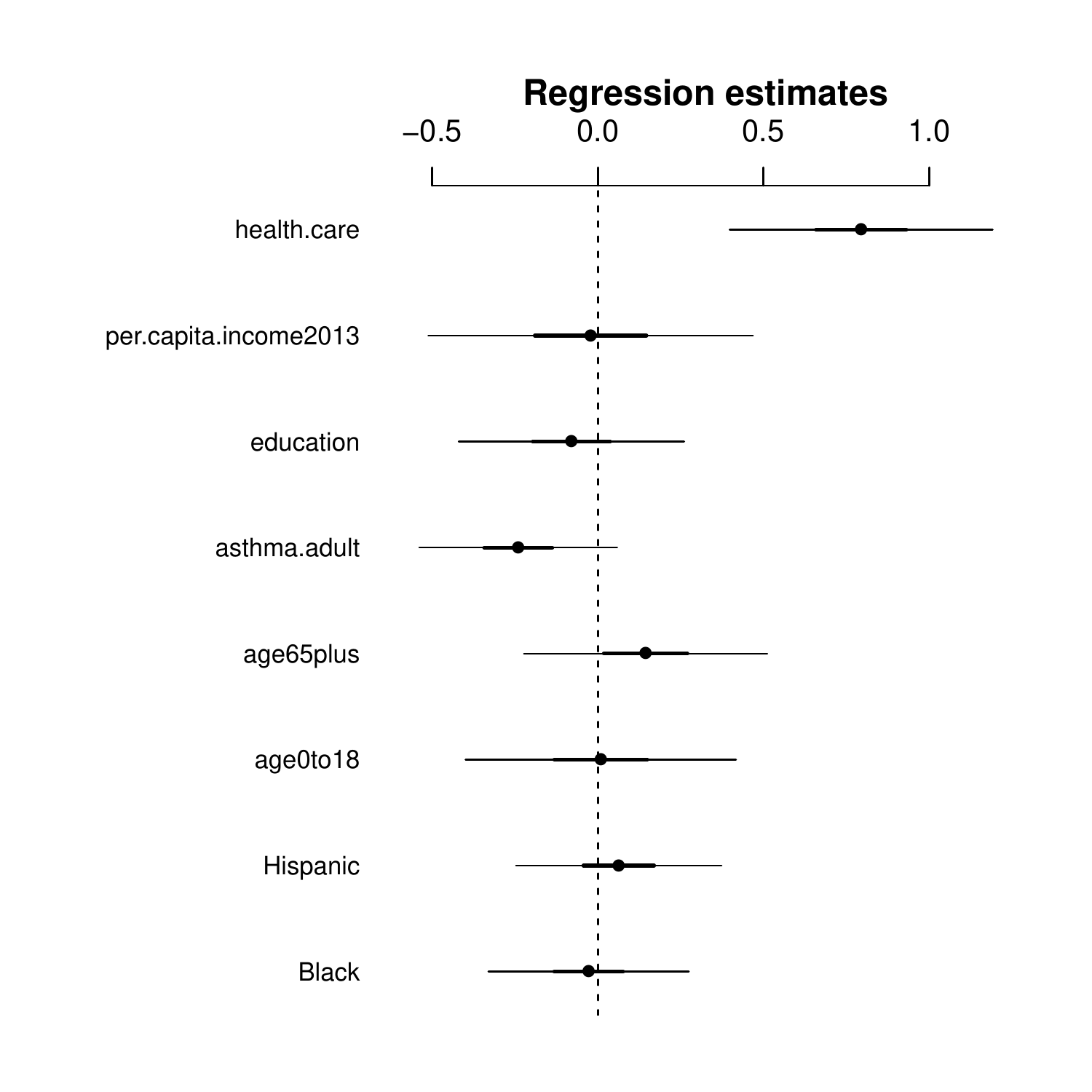}
\caption{Estimates for the overall population without or with region factor.}
\end{figure}

In Figure S1 left panel, we presented the regression estimates of the linear model of the elderly population's mean influenza vaccination coverage rates (Model 3). There were two statistically significant factors: health care access which is positively associated and prevalence of adult asthma is negatively associated with the elderly's vaccination rates.
Other factors positively associated with these rates are educational attainment and the proportion of elderly.
Factors negatively associated are per capita income level, proportion of children, proportion of Hispanic and proportion of Black.

In Figure S1 right panel, we presented the regression estimates of the linear mixed effects model of the elderly population's mean influenza vaccination coverage rates (Model 4). Compared with Model 3, we have included region as a random factor.
Compared with the left panel, Hispanic becomes positively associated with the elderly's vaccination rates.
Other factors showed only non-significant changes in the magnitude of the regression estimates and there were no changes in the direction of the statistical association.

In Figure S2 left panel, we presented the regression estimates of the linear model of the children population's mean influenza vaccination coverage rates (Model 5). Health care access was the only statistically significant factor, and it is positively associated with the children's influenza vaccination coverage rates.
Other factors showing positive associations are: per capita income level, prevalence of asthma among adults, proportion of elderly, proportion of children and the proportion of Hispanic.
Factors showing negative associations are: educational attainment and the proportion of Black.

In Figure S2 right panel, we presented the regression estimates of the linear mixed effects model of the children's population's mean influenza vaccination coverage rates (Model 6). Compared with Model 5, we have included region as a random factor.
Compared with the left panel, proportion of black becomes positively associated with the children's vaccination rates.
Other factors showed only non-significant changes in the magnitude of the regression estimates and there were no changes in the direction of the statistical association.

\section*{Discussion}
Our results showed that health care access is the only factor that is consistently positively and significantly associated with the influenza vaccination coverage rates, both with and without region as a factor, and across the overall, elderly and children's populations.
We also found that the prevalence of asthma in adults is negatively and significantly associated with the elderly's influenza vaccination coverage rates.
For other factors we considered, we noted that per capita income level, proportion of elderly, proportion of children are consistently positively associated with the influenza vaccination coverage rates, while educational attainment are negatively associated with vaccination rates across the six models.

The directional relationship were different across the models for race and ethnic origins, in both the proportion of Hispanic and Black.
Hispanic was positively associated with vaccination rates in five models except for Model 3, the elderly model without region as a factor.
Black was negatively associated with vaccination rates in five models except for Model 1, the children's model with region as a factor.

Our results were consistent with Fox and Shaw \cite{Fox} who showed that at individual level, adults with health care access are more likely to receive influenza vaccination than those without such access.

Bish et al conducted a systematic review and found that being male, of older age, being an ethnic minority are demographic factors associated with uptake of vaccination against pandemic influenza \cite{Bish}. Brien et al conducted another systematic review and found that being male, younger age, higher educational level, being a doctor, being in a priority group for influenza vaccination, receiving a seasonal vaccination in the past, believing in the safety and efficacy of the vaccine, and obtaining vaccine-related information from medical sources, are associated with uptake of pandemic influenza vaccination \cite{Brien}. Comparing with their results, our ecological analyses also showed educational attainment was a significant factor.
These two systematic reviews were different from our study mainly because they considered pandemic influenza vaccination only, and they covered individual-level studies from many countries have varying levels of vaccination coverages.

Previous research studied the racial disparities in influenza vaccination coverage in the U.S. from 2007/08 through 2011/12. Their results showed that at individual level, vaccination coverage rates among the adult subgroups were lower among Hispanics and non-Hispanic blacks, when compared to non-Hispanic whites\cite{Lu2}. Uscher-Pines et al. found that Hispanics and non-Hispanic whites have similar vaccination rates for pandemic influenza, and they suggested this could be due to the Mexican origin of the pandemic that leads to heightened awareness among Hispanics for pandemic influenza vaccination \cite{Uscher-Pines}. They also found that seasonal vaccination in 2009/10 was significantly higher among non-Hispanic whites than non-Hispanic blacks and Hispanics \cite{Uscher-Pines}. We found a consistent, but not significant ecological association. Our results showed that Hispanics were ecologically associated with higher mean vaccination coverage rates in the overall and children populations. These results showed that targeted efforts to eliminate racial disparities are still needed in order to improve the overall vaccination coverage levels.

Strengths of this study include a comparison of ecological factors across six different models, covering overall, elderly and children populations and taking region into account. We were also able to investigate several potentially relevant socio-demographic and health-related factors.

The main study limitation is that it is a state-level ecological study, so we cannot make direct conclusions about causation, nor examine effects at the individual level.
Further, although we have studied a number of potentially relevant factors, other factors might be important at a relevant scale -- for example, influenza epidemic levels, mass media reports on influenza, reimbursement rates for influenza vaccination, and vaccine availablility \cite{Yoo}.

In this study, we have found that neighboring states tend to display similar vaccination coverage rates; future studies could employ more formal techniques to investigate geospatial clustering of vaccination.
We could also look at whether the burdens of influenza-related illnesses and deaths of the earlier years are correlated with the influenza vaccination coverage rates of the current year.

\section*{Conclusions}
We have analysed factors associated with vaccination coverage rates at the state level.
Health care access is the only factor that is positively and significantly associated with mean influenza vaccination coverage rates in all of overall, elderly and children populations, both with and without region as a factor.
We also found that prevalence of asthma among adults are negative and significantly associated with elderly's mean vaccination coverage rates.
Policymakers can make use of this information to identify target populations and to tailor future influenza vaccination campaigns according to these variations, so as to achieve the Healthy People influenza vaccination goals in 2020 for both children and adults \cite{HP}.	


\section*{Acknowledgments}
DH was supported by the Early Career Scheme of the Hong Kong Research Grants Council (PolyU 251001/14M). 

\section*{Author Contributions}

AC, DY, JD and DH conceived and designed the experiments, performed the experiments, analyzed the data, and wrote the paper.

\section*{Conflict of interest}
The authors claim no conflict of interest.

\end{document}